\documentclass[prd,twocolumn,amsmath,amssymb,nofootinbib,preprintnumbers,balancelastpage]{revtex4}

\usepackage[english]{babel}
\usepackage{graphicx}
\usepackage{hyperref}

\allowdisplaybreaks

\hypersetup{
    pdfnewwindow=true,      
    colorlinks=true,        
    linkcolor=black,        
    citecolor=blue,         
    filecolor=blue,         
    urlcolor=blue           
}
\begin{document}
\title{\Large {\bf{Baryonic Dark Matter}}}
\author{Michael Duerr}
\email{michael.duerr@mpi-hd.mpg.de}
\author{Pavel Fileviez P\'erez}
\email{fileviez@mpi-hd.mpg.de \\}
\affiliation{\vspace{0.15cm} \\  Particle and Astro-Particle Physics Division \\
Max Planck Institute for Nuclear Physics {\rm{(MPIK)}} \\
Saupfercheckweg 1, 69117 Heidelberg, Germany}

\begin{abstract}
We investigate a simple extension of the Standard Model where the baryon number is a local gauge symmetry 
and the cold dark matter in the Universe can be described by a fermionic field with baryon number. We refer to this 
scenario as ``Baryonic Dark Matter''. The stability of the dark matter candidate is a natural consequence of the spontaneous 
breaking of baryon number at the low scale and there is no need to impose an extra discrete symmetry. 
The constraints from the relic density and the predictions for direct detection are discussed in detail.
We briefly discuss the testability of this model using the correlation between the Large Hadron Collider data and possible results from dark matter experiments. 
\end{abstract}

\maketitle

\section{Introduction}
The existence of dark matter (DM) in the Universe has motivated many experimental studies and theoretical speculations.
Today we know that about $26 \%$ of the energy density of the Universe is in form of cold dark matter but we have no idea 
about the origin and nature of this type of matter. There are a lot of theoretical ideas to describe the properties of 
the dark matter sector, which can be as complex as the visible sector. Among the very popular candidates 
are the weakly interacting massive particles which appear in several extensions of the Standard Model (SM) of particle physics. 
Thanks to many experimental collaborations, there are relevant constraints on the properties of these 
candidates which play an important role in ruling out some of the theories for dark matter. For a review on 
different dark matter candidates and experiments see Ref.~\cite{Drees}.

We distinguish between baryonic and non-baryonic matter in the Universe, and we say that the cold dark matter is non-baryonic. 
This refers to the fact that it has to be different from the ordinary matter that is made of quarks (and leptons). As is well known the quarks 
are the only particles in the context of the SM that carry baryon number, and they form protons and neutrons. 
In this article we will discuss a different scenario where the dark matter carries also baryon number.   

Baryon number is an accidental global symmetry of the renormalizable couplings of the Standard Model Lagrangian, but we know 
that it has to be broken to explain the matter--antimatter asymmetry of the Universe. Recently, in Ref.~\cite{Duerr:2013dza}, we have 
proposed the simplest realistic model where it is possible to have the spontaneous breaking of the baryon and lepton numbers. 
We will not be concerned with lepton number in this article and we will discuss only the spontaneous breaking of baryon number. 
See also Refs.~\cite{Pais:1973mi,Rajpoot:1987yg,FileviezPerez:2010gw,Dulaney:2010dj,FileviezPerez:2011pt} for earlier attempts to gauge baryon 
and lepton numbers. 

In this article we focus on a simplified version of the model in Ref.~\cite{Duerr:2013dza}, where there is an interesting connection between the 
stability of the dark matter and the spontaneous breaking of the local baryon number. We investigate the properties of a fermionic dark 
matter candidate which has baryon number and refer to this type of scenario as ``Baryonic Dark Matter.''  We show the relic density 
constraints and the predictions for direct detection experiments. Since this model has only four free parameters one could hope to test this idea 
combining the possible results from the Large Hadron Collider and dark matter experiments. Note that we assume the dark matter to be a thermal relic in this article. The asymmetric component of dark matter in this model has been discussed recently in Ref.~\cite{Perez:2013tea}.

This article is organized as follows: In section II we discuss the theoretical framework, while in section III we calculate the DM relic density. In Sec. IV, we show the correlation between the bounds from direct detection experiments and the relic density.
Additionally, we discuss a possible test of this model by combining the efforts at the Large Hadron Collider and dark matter experiments. We summarize and conclude in section V. 
\section{Baryon Number and Dark Matter}
Recently, we have proposed a simple extension of the Standard Model where one can 
understand the spontaneous breaking of baryon and lepton numbers at the low scale~\cite{Duerr:2013dza}.
Here, we will discuss a simplified version of this model, only considering baryon number as a local gauge symmetry.
Therefore, this model is based on the gauge group
 \begin{equation*}
  SU(3)_C \otimes SU(2)_L \otimes U(1)_Y \otimes U(1)_B .
 \end{equation*}
In order to define an anomaly-free theory using this gauge group, we need to include additional fermions that account for anomaly cancellation,
\begin{align} 
\Psi_L  & \sim  (1, 2, -\frac{1}{2}, B_1), & 
\Psi_R &\sim (1, 2, -\frac{1}{2}, B_2), \\
\eta_R &\sim  (1, 1, -1, B_1), &
\eta_L &\sim (1, 1, -1, B_2), \\
\chi_R &\sim (1, 1, 0, B_1), &
\chi_L &\sim (1, 1, 0, B_2),
\end{align}
and extend the scalar sector with a new Higgs boson 
to allow for a spontaneous breaking of baryon number, 
\begin{equation}
S_{B} \sim (1, 1, 0, -3). 
\end{equation}
Here $B_1$ and $B_2$ refer to the baryon numbers of the additional fermions 
which are vector-like under the SM gauge group. From the conditions that 
ensure the cancellation of all relevant baryonic anomalies, one finds the 
 following relation for the baryon numbers of the new fermions:
  \begin{equation}\label{eq:B}
   B_1 - B_2 = -3.
  \end{equation}
The relevant interactions of the new fields in the theory are 
\begin{equation}
-\mathcal{L} \supset \lambda_1 \bar{\Psi}_L \Psi_R S_{B} + \lambda_2 \bar{\eta}_R \eta_L S_{B} + \lambda_3 \bar{\chi}_R \chi_L S_{B} + \text{h.c.}
\end{equation}
Notice that one can have terms such as $a_1 \chi_L \chi_L S_{B}$ and $a_2 \chi_R \chi_R S_{B}^\dagger$ only when $B_1 = -B_2$. Here we will stick 
to the case where $B_1 \neq - B_2$. The Yukawa interactions between the 
new fields and the Standard Model Higgs boson are present as well, but they are not relevant for our main discussion. It is important to notice 
that the new Higgs boson must have baryon number $-3$ in order to generate vector-like mass for the new fermions. Therefore, once $S_B$ gets a vacuum 
expectation value breaking local $U(1)_B$ we never generate any operator mediating proton decay and the scale for baryon number violation can be as low 
as a few TeV (or even below). For a discussion of the bounds on the mass of a leptophobic neutral gauge boson see Refs.~\cite{Dobrescu:2013cmh,An:2012ue,Frandsen:2012rk}.
    
In this simple theory, when the local baryon number is spontaneously broken by the vacuum expectation value $v_B$ of $S_B$, one obtains 
a $Z_2$ discrete symmetry which protects the stability of the dark matter candidate. Under this $Z_2$ the new fermionic fields transform as 
\begin{equation*}
\Psi_{L,R} \to - \Psi_{L,R}, \  \eta_{L,R} \to - \eta_{L,R}, {\rm{and}} \ \chi_{L,R} \to - \chi_{L,R}. 
\end{equation*}
Therefore, when the lightest new field with baryon number is neutral, one can have a candidate for the cold dark matter in the Universe. 
It is important to mention that the main idea of having a dark matter candidate with baryon number was discussed earlier in 
Refs.~\cite{Agashe:2004ci,Farrar:2005zd,FileviezPerez:2010gw}.

For simplicity, we will focus on the case when the dark matter is SM singlet-like and is the Dirac fermion  $\chi = \chi_L + \chi_R$. Since the dark matter has baryon number, the relevant interactions with the new gauge boson $Z_B$ related to baryon number are
\begin{equation}
\mathcal{L} \supset g_B \bar{\chi} \gamma_\mu  Z_B^\mu \left( B_2 P_L + B_1 P_R \right) \chi,
\end{equation}
where $P_L$ and $P_R$ are the left- and right-handed projectors, and $g_B$ is the gauge coupling related to baryon number.
Of course, the new gauge boson also couples to the SM quarks, which is crucial to understand the properties of the dark matter candidate.
The leptophobic gauge boson mass reads as
\begin{equation}
M_{Z_B}=3 g_B v_B,
\end{equation}
while the mass of the SM singlet-like baryonic dark matter candidate is given by
\begin{equation}
M_\chi= \lambda_3 v_B / \sqrt{2} < \frac{\sqrt{2 \pi}}{3} \frac{M_{Z_B}}{g_B}.
\end{equation}
This upper limit is coming from the perturbative condition on the Yukawa coupling $\lambda_3$, i.e. $|\lambda_3|^2/4 \pi <1$.

It is important to notice that this model for baryonic dark matter has only four free parameters:
$$M_\chi, \ M_{Z_B}, \ g_B, \ {\rm{and}} \ B,$$
and one needs to satisfy the relic density constraints and the bounds from direct detection.
Here $B=B_1 + B_2$ is the parameter which enters in the predictions for the relevant cross sections.
One could imagine that the parameters $M_{Z_B}$ and $g_B$ can be determined from the discovery 
of a leptophobic gauge boson at the Large Hadron Collider. Therefore, one can say that for a given value 
of these two quantities we can find the values of $B$ and $M_\chi$ using the relic density and 
spin-independent cross section values. We will discuss in more details the numerical predictions 
in the following sections in order to appreciate this connection between collider physics and dark matter experiments.
In the rest of the paper we will neglect the kinetic mixing between $U(1)_B$ and $U(1)_Y$, as well as the mixing between the SM Higgs 
and $S_B$.    
%
\section{Dark Matter Relic Density}
The dark matter particle $\chi$ can annihilate into two standard model quarks through the interaction with the leptophobic gauge boson $Z_B$. The annihilation cross section is given by
\begin{equation}
 \sigma v = \sum_q \frac{g_B^4 \sqrt{1- 4 M_q^2/s} }{72 \pi s^{1/2} [ ( s - M_{Z_B}^2)^2 + \Gamma_{Z_B}^2 M_{Z_B}^2]} \frac{C_B}{M_\chi},  \\
\end{equation}
with
\begin{eqnarray} 
C_B &=& \left(B_1^2 + B_2^2\right) \left(  s^2 + s (2 M_q^2 - M_\chi^2) - 2 M_\chi^2 M_q^2 \right) \nonumber \\
&+ & 6 B_1 B_2 M_\chi^2 (s + 2 M_q^2).
\end{eqnarray}
Here $M_q$ is the mass of the quarks, $s$ is the square of the center-of-mass energy and $\Gamma_{Z_B}$ is the decay width of the leptophobic gauge boson. 
In the non-relativistic limit, the above cross section reads as
\begin{equation}
\sigma v  \approx  \sum_q \frac{B^2 g_B^4 (2 M_\chi^2 + M_q^2) \sqrt{1- M_q^2 / M_\chi^2}}{24 (M_{Z_B}^2 -4 M_\chi^2)^2 \pi} \equiv \sigma_0.
\end{equation}
We neglected the decay width of the new gauge boson for simplicity. Note that for $B_1 = -B_2$, which would allow for terms leading to Majorana masses for the DM fields after symmetry breaking, $\sigma_0 = 0$ and the annihilation cross section is velocity suppressed. 

\begin{figure}[t]
 \centering
 \includegraphics[width=.98\linewidth]{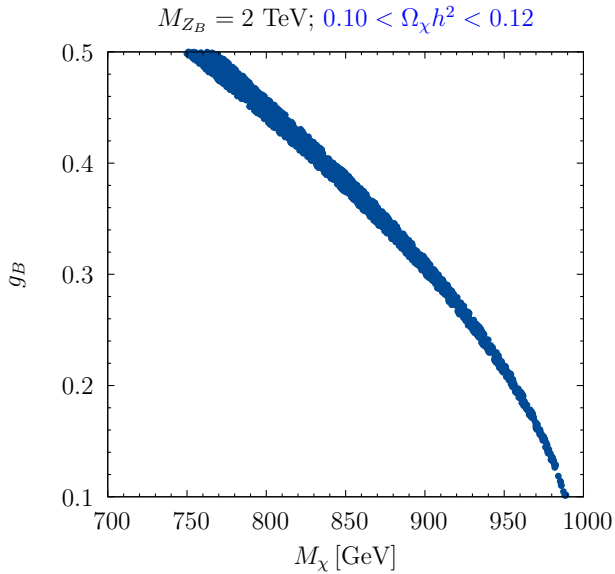}
 \caption{Allowed values for the gauge coupling $g_B$ vs.\ the dark matter mass $M_\chi$ when the DM relic density is in the range 
 $0.10 < \Omega_\chi h^2 < 0.12$.  Here we use $B=1$ and $M_{Z_B}=2$ TeV.
 \label{fig:figure1}}
\end{figure}

As is well known the cold dark matter relic density can be approximated by
\begin{equation}
\Omega_\chi h^2 = \frac{1.07\times 10^9}{\text{GeV}} \left( \frac{x_f}{\sqrt{g_\ast} \sigma_0 M_\text{Pl}} \right), 
\end{equation}
where $x_f = M_\chi / T_f$ is the freeze-out temperature and $M_{\text{Pl}} = 1.22 \times 10^{19}$~GeV is the Planck mass scale. The quantity $x_f$ 
can be calculated using the expression
\begin{eqnarray}
x_f &=& \ln \left[ 0.038 \left( \frac{g}{\sqrt{g_\ast}} \right) M_\text{Pl} M_\chi \sigma_0 \right] \nonumber \\
&-& \frac{1}{2} \ln \left\{ \ln \left[ 0.038 \left( \frac{g}{\sqrt{g_\ast}} \right) M_\text{Pl} M_\chi \sigma_0 \right] \right\},
\end{eqnarray}
where $g$ is the number of internal degrees of freedom, and $g_\ast$ is the effective number of relativistic degrees of freedom evaluated 
around the freeze-out temperature $x_f$. The current value of the DM relic density provided by Planck 
is $\Omega_\text{DM} h^2 = 0.1199 \pm 0.0027$~\cite{Ade:2013zuv}, which we will use to understand the constraints on our model.

As we have mentioned above the relevant parameters for our study are the dark matter mass $M_\chi$, the mass of the new 
gauge boson $M_{Z_B}$, the gauge coupling $g_B$, and the baryon numbers of the new fermionic fields. In order 
to illustrate the numerical results we will choose $B=1$ for simplicity and later discuss how one can test the model for any value of B. 

In Fig.~\ref{fig:figure1} we show the allowed region where the DM relic density is $0.10 < \Omega_\chi h^2 < 0.12$ in the 
plane spanned by the DM mass $M_\chi$ and the gauge coupling $g_B$. For a range of $0.1 < g_B < 0.5$, the DM mass in 
the range $750~\text{GeV} < M_\chi < 990~\text{GeV}$ allows for a DM relic density around the current value. 
Notice that a gauge coupling $g_B$ in the interval $[0.1,0.5]$ and $M_{Z_B}=2$ TeV is consistent with recent 
collider studies~\cite{Dobrescu:2013cmh,TheATLAScollaboration:2013kha}. One could use even smaller values for the gauge boson mass because the 
collider bounds are not so strong for this type of gauge bosons.

\begin{figure}[t]
 \centering
 \includegraphics[width=1\linewidth]{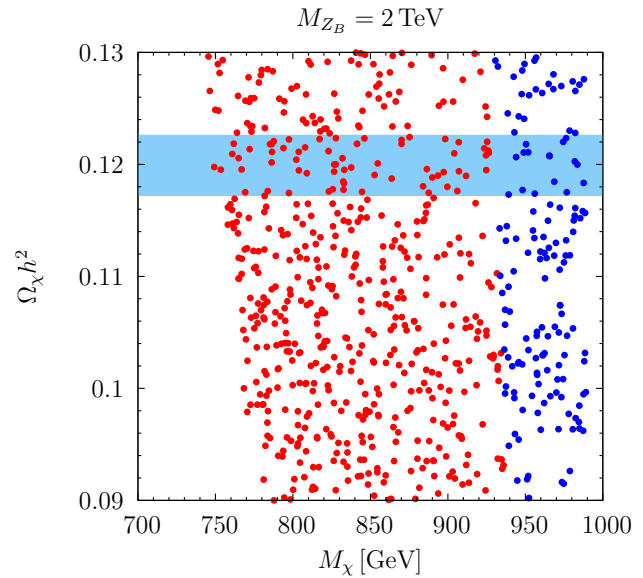}
 \caption{Values of the DM relic density $\Omega_\chi h^2$ vs.\ the dark matter mass $M_\chi$. The gauge boson mass was chosen to be $M_{Z_B}  = 2~\text{TeV}$. Blue dots are for values of $g_B \in [0.10,0.25]$, red dots are for values of $g_B \in [0.25,0.50]$. The blue band shows the allowed range for the DM relic density.  \label{fig:figure2}}
\end{figure}

In order to illustrate the possible values of the relic density for different values of the free parameters, Fig.~\ref{fig:figure2} shows possible values of the DM relic density for DM masses $M_\chi \in [700,1000]~\text{GeV}$ for $M_{Z_B} = 2~\text{TeV}$. The blue dots correspond to the solutions when $g_B \in [0.10,0.25]$, and the red dots are for values 
of $g_B \in [0.25,0.50]$. The region of the current relic density measured by Planck~\cite{Ade:2013zuv} is marked by a blue band. In this band, many solutions exist. Of course, our DM candidate could make up only part of the total DM relic density, and many solutions for  $\Omega_\chi h^2$ smaller than the current value also exist. As one can appreciate from these numerical results, there is no need to be on the resonance to achieve the correct relic density value and all the solutions can be in agreement with the collider constraints. In our opinion, the simplicity of this model is very appealing and one can make predictions for direct detection as well, which we discuss in the next section.
\section{Direct Detection}
%
The direct detection of the baryonic dark matter candidate is also through the baryonic force. 
The elastic spin-independent nucleon--dark matter cross section is given by
\begin{equation}
\sigma_{\chi N}^{\text{SI}}=\frac{M_N^2 M_\chi^2}{4 \pi (M_N + M_\chi)^2} \frac{g_B^4}{M_{Z_B}^4} B^2.
\end{equation}
Notice that the numerical results will be independent of the matrix elements because the baryon number is a conserved current in the theory.
This is a nice feature of the model because we do not introduce any extra unknown parameter coming from the matrix elements, as one 
has in several dark matter models such as the Higgs portal. 

In order to show our numerical results we can write the above expression as
\begin{equation}
\sigma_{\chi N}^{\text{SI}} ({\rm{cm}}^2) = 3.1 \times 10^{-41} \  \left(\frac{\mu}{1~\text{GeV}}\right)^2 \ \left(\frac{1 \ {\rm{TeV}}}{r_{B}}\right)^4 B^2 \ {\rm{cm}}^2,
\end{equation}
where $\mu=M_N M_\chi/(M_N + M_\chi)$ is the reduced mass and $r_B=M_{Z_B}/g_B$. 

\begin{figure*}
 \centering
 \includegraphics[width=.7\linewidth]{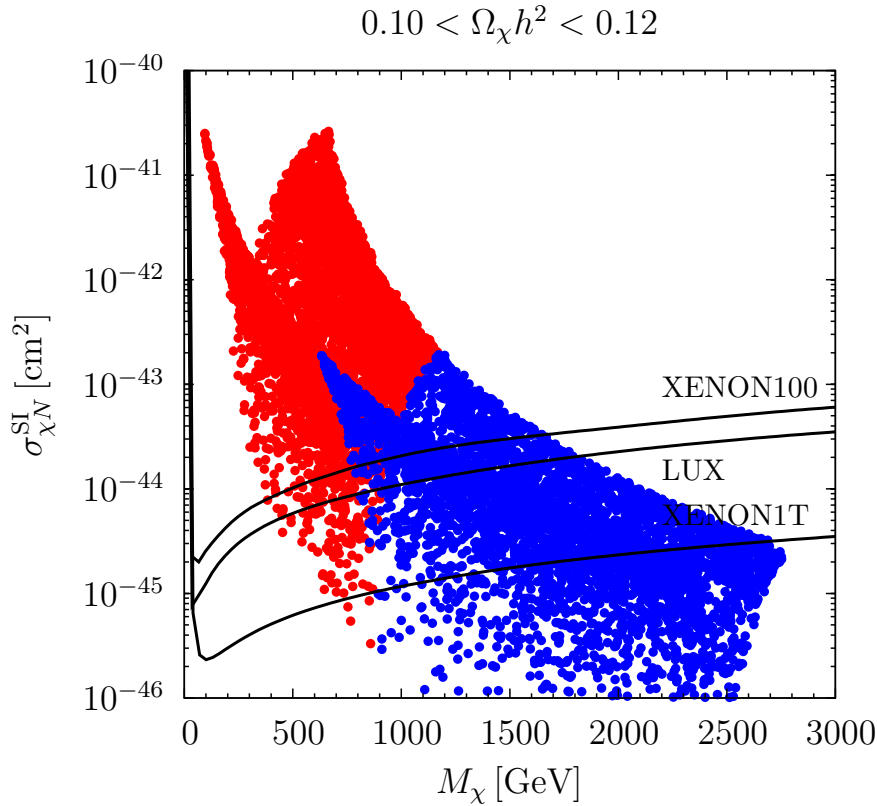}
 \caption{Spin-independent elastic DM--nucleon cross section $\sigma_{\chi N}^{\text{SI}}$ as a function of the baryonic dark matter mass $M_\chi$. The exclusion limits 
 of XENON100 and LUX are shown, as well as the projected values for XENON1T. Red dots are for values $M_{Z_B} \in [500~\text{GeV},1700~\text{GeV}]$, blue dots are for values $M_{Z_B} \in [1700~\text{GeV},5000~\text{GeV}]$.\label{fig:figure3}}
\end{figure*}

The numerical results for the spin-independent elastic DM--nucleon cross section are shown in Fig.~\ref{fig:figure3} 
as a function of the dark matter mass $M_\chi$ for two ranges of $M_{Z_B}$. For definiteness, we choose the same value $B=1$ as in the analysis 
for the relic density changing $g_B$ between 0.1 and 0.5. 
One can appreciate that the XENON100~\cite{Aprile:2012nq} and the LUX~\cite{Akerib:2013tjd} bounds on $\sigma_{\chi N}^{\text{SI}}$ significantly cut into the 
parameter space and allow for DM masses larger than about 400 GeV. The projected values for XENON1T~\cite{Aprile:2012zx} even 
constrain the range of dark matter masses more tightly, and allow for values larger than 700 GeV. 
It is important to note that the range $M_Z > 1.7~\text{TeV}$ allowed by current LHC results~\cite{TheATLAScollaboration:2013kha} (see blue dots in Fig.~3) gives several scenarios consistent with the relic density constraints and the XENON100 and LUX bounds. We have to say that thanks to the experimental collaborations we can rule out a large fraction 
of the parameter space and with future experiments such as XENON1T we will be able to rule out most of the solutions for the dark matter mass below 1 TeV 
in case of no discovery.

Let us discuss the possible correlation between possible discoveries at the Large Hadron Collider and in dark matter experiments. 
At the Large Hadron Collider we could discover the new neutral gauge boson associated to the breaking of the local baryon 
number, the gauge boson $Z_B$. Therefore, one could know about the mass $M_{Z_B}$ and the gauge coupling $g_B$.  
Assuming that our dark matter candidate describes all the relic density in the Universe and for a given value of the spin-independent 
cross section one can solve for the dark matter mass $M_\chi$ and the baryon number $B$. Then, we could predict 
the values for the production cross section of a dark matter pair and an energetic jet, relevant for the monojet 
searches at the Large Hadron Collider. A possible benchmark scenario is when $g_B=0.2$, $M_{Z_B}=2$ TeV, $M_\chi =955$ GeV, 
and $\sigma_{\chi N}^{\text{SI}} \approx 3.1 \times 10^{-45} \ {\rm{cm}^2}$.  
In summary, one could say that this theory provides a scenario for 
dark matter which could be fully tested in the future combining dark matter and collider experiments.
\section{Summary and Outlook}
We have proposed a simple theory for dark matter where the cold dark matter candidate has spin one-half 
and carries baryon number. We refer to this type of dark matter scenario as ``Baryonic Dark Matter.'' The 
baryon number is defined as a local gauge symmetry which is spontaneously broken 
at the low scale and the stability of the dark matter is a natural consequence coming from symmetry breaking.
This theory for dark matter has only four free parameters which determine the relic 
density and predictions for the spin-independent cross section relevant for direct detection experiments.  

We have shown several numerical results in order to illustrate the possibility to have a consistent scenario 
for cosmology in agreement with the bounds from dark matter experiments. In particular, a nice feature of this setup is to have a viable region of dark matter masses in the TeV-range, which is therefore also testable at the LHC. All in all, this theory provides a scenario for dark matter which could be fully tested in the future 
combining dark matter and collider experiments. In a future publication we will investigate 
the possibility to test this theory at the Large Hadron Collider and the predictions for indirect detection experiments.

\vspace{2ex}

{\textit{Acknowledgments:}}
{\small {P.F.P.\ thanks Mark Brian Wise for discussions.
M.D. is supported by the International Max Planck Research School for Precision Tests of Fundamental Symmetries. }}



\end{document}